\documentclass{emulateapj}

\newcommand{\ergps}{erg\thinspace s$^{-1}$}
\newcommand{\psqcm}{cm$^{-2}$}
\newcommand{\nH}{$N_{\rm H}$}
\newcommand{\Ms}{$M_{\odot}$}
\newcommand{\ts}{\thinspace} 

\slugcomment{Accepted for publication in ApJ Letters}

\shorttitle{Fe XXV from the GOALS LIRGs}
\shortauthors{Iwasawa et al.}

\begin{document}

\title{High-ionization Fe K emission from luminous infrared galaxies}

\author{K. Iwasawa\altaffilmark{1}}

\author{D.B Sanders\altaffilmark{2}}

\author{A.S. Evans\altaffilmark{3}}

\author{J.M. Mazzarella\altaffilmark{4}}

\author{L. Armus\altaffilmark{5}}

\author{J.A. Surace\altaffilmark{5}}

\altaffiltext{1}{INAF-Ossservatorio Astronomico di Bologna, Via Ranzani, 1, 40127 Bologna, Italy}

\altaffiltext{2}{Institute for Astronomy, 2680 Woodlawn Drive, Honolulu, Hawaii 96822-1839}

\altaffiltext{3}{Department of Astronomy, University of Virginia, 530 McCormick Road, Charlottesville, VA 22904 and NRAO, 520 Edgemont Road, Charlottesville, VA 22903-2475}

\altaffiltext{4}{IPAC, California Institute of Technology, Pasadena, CA 91125}

\altaffiltext{5}{{\it Spitzer} Science Center, California Institute of Technology, Pasadena, CA 91125}

\begin{abstract} 
  The Chandra component of the Great Observatories All-Sky LIRG Survey
  (GOALS) presently contains 44 luminous and ultraluminous infrared
  galaxies with log ($L_{\rm IR}/L_{\odot}) = 11.73-12.57$. Omitting
  15 obvious AGNs, the other galaxies are, on average, underluminous
  in the 2-10{\ts}keV band by 0.7{\ts}dex at a given far-infrared
  luminosity, compared to nearby star-forming galaxies with lower star
  formation rates. The integrated spectrum of these hard X-ray quiet
  galaxies shows strong high-ionization Fe K emission (Fe{\ts}{\sc
    xxv} at 6.7{\ts}keV), which is incompatible with X-ray binaries as
  its origin. The X-ray quietness and the Fe{\ts}K feature could be
  explained by hot gas produced in a starburst, provided that the
  accompanying copious emission from high-mass X-ray binaries is
  somehow suppressed. Alternatively, these galaxies may contain deeply
  embedded supermassive black holes that power the bulk of their
  infrared luminosity and only faint photoionized gas is visible, as
  seen in some ULIRGs with Compton-thick AGN.
\end{abstract}


\section{introduction}

While far-infrared (FIR) luminosity\footnote{$L_{\rm FIR} \equiv L
  (40-400\mu m)$, as determined using the prescription described in
  Lonsdale et al. (1985)} is often assumed to be a good indicator of
the star formation rate (SFR) in extragalactic objects (e.g. Young \&
Scoville 1991, Kennicutt 1998), it has also been argued that, in the
absence of active galactic nuclei (AGN), X-ray emission could trace
the SFR as well; this latter assumption is based primarily on the
observed correlation between X-ray luminosity (usually in the
2-10{\ts}keV energy range) and FIR luminosity initially found for
nearby actively star-forming galaxies with SFRs $\sim $3-30 \Ms
yr$^{-1}$ (e.g., Ranalli et al.  2003, Grimm et al. 2003; Gilfanov et
al. 2004). This relation appears to extend to objects with higher SFR
(up to 100 \Ms yr$^{-1}$) and higher redshift (e.g., Hornschemeier et
al. 2005; Persic \& Raphaeli 2007, Lehmer et al. 2008). The hard X-ray
emission is generally assumed to have a power-law form and to be
predominantly due to the collective emission of high-mass X-ray
binaries (HMXBs), which are accreting compact objects (neutron stars
or black holes) formed following the death of short-lived massive
stars, and therefore expected to have a close relationship to the
starburst.

There have been, however, indications that objects with very high
computed SFRs, such as ultraluminous infrared galaxies (ULIRGs), might
depart from the correlation because they are underluminous in X-rays
for their computed SFRs (Persic \& Raphaeli 2007, Barger, Cowie \&
Wang 2007). Arp~220, the nearest ULIRG, is a prime example, with a
2-10{\ts}keV luminosity $\sim 1${\ts}dex below the general $L_{\rm
  X}$-$L_{\rm FIR}$ correlation (see Iwasawa et al. 2005). A further
puzzle comes from its X-ray spectrum, which shows a strong
high-ionization Fe{\ts}K line (mainly Fe{\ts}{\sc xxv}, Iwasawa et
al. 2005). This latter finding means that Arp~220 is not only X-ray
under-luminous, but that the hard X-ray emission is not primarily due
to X-ray binaries because of the presence of a high-ionization
Fe{\ts}K line. In nearby star-forming galaxies, Fe{\ts}{\sc xxv} is much weaker
(e.g., Cappi et al 1999) or undetected, which is consistent that the
hard X-ray emission is dominated by HMXBs. These peculiar properties
of the nearest ULIRGs clearly warrant further investigation with a
larger sample of (U)LIRGs.

The Great Observatory All-sky LIRGs Survey (GOALS)\footnote{More
  information about GOALS is available at
  http://goals.ipac.caltech.edu/} is a multi-wavelength study of the
most luminous infrared-selected galaxies in the local Universe,
selected from the flux-limited {\it IRAS} Revised Bright Galaxy Sample
(RBGS: Sanders et al. 2003). An overview of GOALS is given in Armus
et al. (2009). C-GOALS (PI: D. B. Sanders) is the X-ray component
of the project utilizing data from the Chandra X-ray Observaotry ({\it
  Chandra}, hereafter).  Details of the X-ray observations are
described in Iwasawa et al. in prep.  Here we report results focusing
primarily on the $L_{\rm X}$- $L_{\rm FIR}$ relation and the spectral properties
of the hard X-ray emission, with special attention to the Fe{\ts}K
band.

The cosmology used to calculate luminosities in this paper is
$H_0=70$ km s$^{-1}$ Mpc $^{-1}$, $\Omega_{\rm M}=0.3$,
$\Omega_{\Lambda} =0.7$, based on the latest WMAP results 
(Hinshaw et al. 2009).

\section{The Sample}

The current C-GOALS sample is complete down to log{\ts}$(L_{\rm
  IR}/L_{\odot}) = 11.73$, and consists of 44 galaxies from the RBGS
with redshifts $z = 0.010-0.088$ (see Table~1). We first removed
obvious AGN as follows. The primary criterion was a flat X-ray
spectrum, assessed by the X-ray color or hardness ratio (HR). The
X-ray color is defined as $HR = (H-S)/(H+S)$, where $H$ is the
2-8{\ts}keV counts and $S$ is the 0.5-2{\ts}keV counts. Objects with
$HR > -0.3$ are classified as an AGN. This threshold is chosen because
ULIRGs known to host AGN (Mrk 231, Mrk 273, UGC 5101) cluster just
above this value. All of the optically identified AGN are selected by
this criterion. However, Compton-thick AGN are generally missed by
this criterion because of their weakness in the hard band. Therefore,
objects that show a strong Fe{\ts}K line at 6.4{\ts}keV, a
characteristic signature of a Compton-thick AGN, are also classified
as AGN (NGC 6240, NGC 3690 West, VV 340a). These criteria classify 15
objects as AGN, and they are excluded from further discussion leaving
an ``hard X-ray quiet'' (HXQ, as defined by their small HR) sample of
30 objects (including NGC 3690 East). The SFR
of these 30 HXQ galaxies, calculated assuming that their FIR
luminosity is due to dust heated by star formation alone, ranges from
60 to 300 $M_{\odot}$ yr$^{-1}$.

\begin{table}
\setlength{\tabcolsep}{0.015in}
\begin{center}
  \caption{The C-GOALS sample.}
\begin{tabular}{lcrlcr}
Object & log$L_{\rm FIR}$ & log$L_{\rm HX}$ & \hspace{2mm}Object & log$L_{\rm FIR}$ & log$L_{\rm HX}$ \\
& $L_{\odot}$ & erg s$^{-1}$ & & $L_{\odot}$ & erg s$^{-1}$ \\[5pt]
\multicolumn{6}{l}{\bf HXQ sample} \\
F17207--0014 & 12.42 & 41.34 & \hspace{2mm}NGC 3690 E & 11.49 & 41.00 \\
F19297--0406 & 12.39 & 41.26 & \hspace{2mm}ESO 593-IG8 & 11.83 & 41.28 \\
P07251--0248 & 12.34 & $<40.90$ & \hspace{2mm}VV 705 & 11.79 & 40.83 \\
F12112+0305 & 12.30 & 41.60 & \hspace{2mm}ESO 255-IG7 & 11.75 & 41.46 \\
Arp 220 & 12.22 & 40.96 & \hspace{2mm}F18293--3413 & 11.76 & 41.15 \\
F22491--1808 & 12.18 & 40.78 & \hspace{2mm}F10173+0828 & 11.76 & 39.82 \\
F23365+3604 & 12.11 & 41.20 & \hspace{2mm}ESO 203-IG1 & 11.84 & $<40.48$ \\
F10565+2448 & 11.99 & 41.20 & \hspace{2mm}F01364--1042 & 11.82 & 41.18 \\
F15250+3608 & 11.98 & $<40.65$ & \hspace{2mm}ESO 239-IG2 & 11.73 & 40.99 \\ 
F09111--1007 & 11.97 & 41.11 & \hspace{2mm}P21101+5810 & 11.71 & 40.30 \\ 
ESO 286-IG19 & 11.98 & 41.32 & \hspace{2mm}VV 250 & 11.68 & 41.49 \\
VII Zw31 & 11.89 & 41.43 & \hspace{2mm}F10038-3338 & 11.69 & 40.80 \\
ESO 69-IG6 & 11.88 & 41.18 & \hspace{2mm}ESO 77-IG14 & 11.67 & 41.15 \\
F17132+5313 & 11.84 & 40.93 & \hspace{2mm}UGC 4881 & 11.65 & 40.73 \\ 
II Zw96 & 11.94 & 41.18 &  \hspace{2mm}IC 883 & 11.65 & 40.81 \\[5pt]
\multicolumn{6}{l}{\bf AGN sample} \\
Mrk 231 & 12.37 & 42.48 & \hspace{2mm}P19542+1110 & 12.06 & 42.61 \\
F14348--1447 & 12.35 & 41.77 & \hspace{2mm}ESO 148-IG2 & 11.95 & 41.92 \\  
P09022--3615 & 12.24 & 42.30 & \hspace{2mm}UGC 5101 & 11.93 & 41.67 \\
P13120--5453 & 12.24 & 41.67 & \hspace{2mm}NGC 3690 W & 11.32 & 41.00 \\
F14378--3651 & 12.16 & 41.53 & \hspace{2mm}NGC 6240 & 11.81 & 42.54 \\
Mrk 273 & 12.15 & 42.40 & \hspace{2mm}ESO 60-IG16 & 11.71 & 41.86 \\
F05189--2524 & 12.08 & 43.11 & \hspace{2mm}VV 340a & 11.66 & 41.46\\                        
F08572+3915 & 12.07 & 41.30 & & \\  
\end{tabular}
\end{center}
\tablecomments{Source names beginning with "F" or
"P" are from the IRAS Faint Source Catalog or Point
Source Catalog, respectively.}
\end{table}

\section{L(2-10{\ts}keV) and L(FIR) relation}

The sensitivity of {\it Chandra} declines steeply above 7{\ts}keV. The
2-10{\ts}keV luminosity is estimated by extrapolating the spectral
model that describes the data up to 7{\ts}keV. When multiple hard
X-ray sources are present in a single object, those which have no {\it
  Spitzer}-MIPS 24 $\mu$m counterpart are excluded for calculating the
2-10{\ts}keV luminosity, assuming they have no relation with the IRAS
measured luminosity. In NGC 3690, the western and eastern galaxies are
treated separately due to the difference in classification, and their
FIR luminosity ratio is assumed to be 1:2 based on the 38 $\mu$m study
(Charmandaris et al. 2002). The X-ray luminosity is as observed in
the rest-frame 2-10{\ts}keV band, corrected only for Galactic
absorption. The median value of the 2-10{\ts}keV luminosity is
$1.4\times 10^{41}$ \ergps (which also matches the mean).

Figure~1 shows a plot of hard X-ray luminosity, $L$(2-10{\ts}keV),
versus FIR luminosity, $L(40-400 \mu{\rm m})$, for our complete sample
of 44 RBGS galaxies listed in Table~1. For the GOALS HXQ galaxies
alone, no clear correlation is seen (Kendal's $\tau\simeq 0.3$) in the
limited FIR luminosity range. The correlation obtained by Ranalli et
al. (2003) is re-derived in the same way as done for our sample: log
$L{\rm (2-10keV)} = -3.84$ log $L{\rm (FIR)}$. This is fully
consistent with the original relation with an offset due to the wider
FIR wavelength range. The $L_{\rm X}$-SFR correlations studied by
various authors (e.g., Grimm et al. 2003, Persic \& Raphaeli 2007,
Lehmer et al. 2008) all lie within a factor of $\pm$2 of Ranalli et
al. (2003) as indicated in the figure. The correlation line is
consistent with the upper envelope of the HXQ sample. Most of our HXQ
galaxies lie well below this correlation. The median value of the
logarithmic ratio of $L$(2-10{\ts}keV) and $L$(FIR) is $-4.5$, which
means our sample of HXQ galaxies is $\sim$0.7{\ts}dex underluminous in
the 2-10{\ts}keV band for a given $L$(FIR), or corresponding SFR. We
note that, given the lack of correlation in our sample, this is merely
a comparison of the average value relative to the $L_{\rm X}$-$L_{\rm FIR}$
correlation.

\begin{figure}
\includegraphics[width=0.37\textwidth,angle=270]{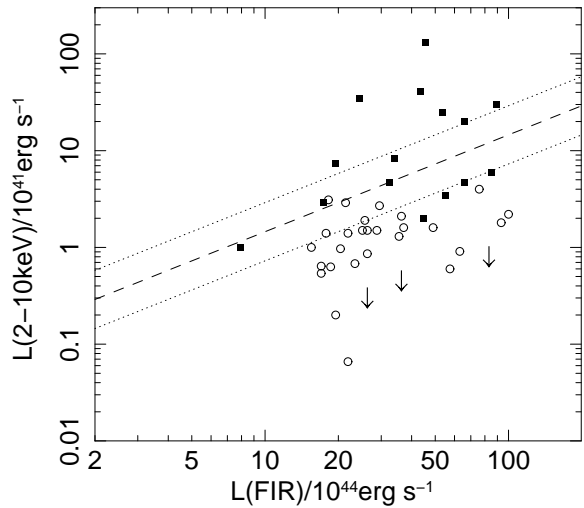}
\caption{$L$(2-10{\ts}keV) versus $L$(FIR) for the GOALS AGN (filled
  squares) and HXQ (open circles) galaxies listed in Table~1. For the
  three objects with no 2-10 keV detection, the 95 per cent upper
  limits derived by following Kraft et al. (1991) are shown. The
  dashed line indicates the correlation for the sample of Ranalli et
  al. (2003). The dotted lines show a factor of 2 above and below the
  Ranalli et al. correlation, within which the other studies of the
  same relationship lie.}
\end{figure}

\section{The integrated 4-8 keV spectrum}

\begin{figure}
\includegraphics[width=0.35\textwidth,angle=270]{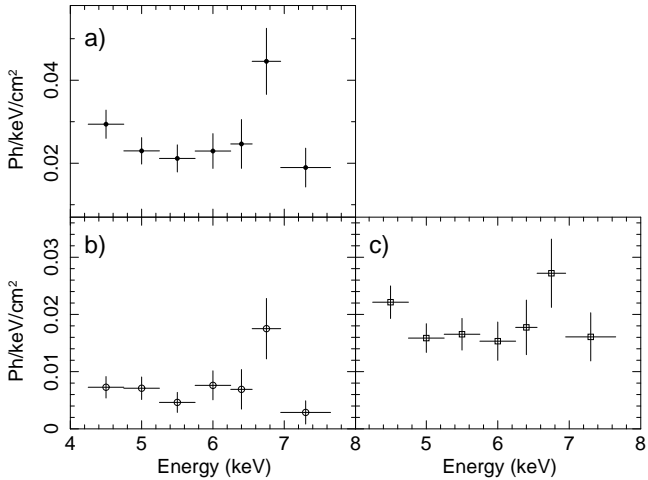}
\caption{ (a) The integrated 4.25-7.65{\ts}keV spectrum of the 29 HXQ
  objects in Table~1. The energy is in the rest-frame. An excess in
  the high-ionization Fe{\ts}K band (primarily due to Fe {\sc xxv}) at
  6.55-6.95 keV is clearly visible. The continuum is flat ($\Gamma =
  1.1\pm 0.5$) and the Fe{\ts}K band excess has $EW\sim 1$ keV.\ (b)
  The integrated spectrum of Arp~220 and NGC~3690{\ts}E, two Fe {\sc
    xxv} emitters known from XMM-Newton observations. c)
  The integrated spectrum of the HXQ sample excluding Arp 220 and NGC
  3690 E. Even without the two known Fe {\sc xxv} emitters, a
  comparable excess flux above the continuum in the Fe {\sc xxv} band
  is present, indicating that other galaxies make a significant
  contribution to the total line flux in a).}
\end{figure}

We investigate the integrated hard-band spectrum of the 30 HXQ
galaxies, since they are too faint in hard X-rays to allow an
individual inspection for the iron line. Although detection of the
Fe{\ts}K line has been reported for the XMM-Newton spectra of NGC
3690{\ts}East (Ballo et al. 2004) and Arp 220 (Iwasawa et al. 2005),
the line does not have sufficient counts in the Chandra data for a
significant detection (see e.g., Clements et al. 2002). These two
objects are included in the integrated hard-band spectrum and their
contribution to the total spectrum is also investigated. In cases
where multiple hard X-ray sources are detected in a single object
(IRAS F09111--1007, ESO 255-IG007, VV 250, ESO 77-IG014, IRAS
F12112+0305, VV 705, UGC 4881), multiple apertures were selected
accordingly.

Given that our primary interest is in the Fe{\ts}K properties, we
restrict the analysis to the Fe{\ts}K and neighboring band
(4.25-7.65{\ts}keV in the rest-frame). The Fe{\ts}K complex generally
consists of two components with distinct ionization states:\ (a) cold
Fe at 6.4{\ts}keV; and\ (b) highly ionized Fe{\ts}{\sc xxv}
(6.7{\ts}keV) with a minor contribution from Fe{\ts}{\sc xxvi}
(7.0{\ts}keV). These two components can be separated well at the
spectral resolution of the Chandra ACIS-S. We define two spectral
bands centered on the respective line components (6.35-6.55{\ts}keV
and 6.55-6.95{\ts}keV), and five neighboring bands for the continuum,
giving seven spectral bands over the range 4.25-7.65{\ts}keV in the
rest-frame.

The hard X-ray sources of our sample are generally compact, and a
small extraction radius, typically $\sim 1.5-2^{\prime\prime}$, is used. The
detected counts have been corrected for background using the data from
a source-free area on the detector in the same observation, although the
correction is almost negligible with such a small aperture.

For individual objects, source counts recorded in seven bands
corresponding to the pre-defined rest-frame bands were
accumulated. The detector response curve was corrected by dividing by
the mean effective area of each spectral band. As most sources have
only a few counts over the energy range, stacking was done as a
straight integration of individual sources, neither normalizing by the
exposure time nor source brightness.

The total source counts accumulated in the rest-frame 4.25-7.65{\ts}keV
band for the HXQ sample are 296{\ts}ct. The integrated spectrum
is shown in Fig.~2a. A strong Fe{\ts}K line is immediately recognized with
a $3\sigma$ excess.  The line peaks at the high-ionization Fe{\ts}K band
(6.55-6.95 keV), indicating that the line is primarily due to Fe{\ts}{\sc xxv}
from a highly ionized medium. The total counts collected in this line
band is 35{\ts}ct. There might also be a slight excess in the 6.4{\ts}keV line
band, which is however not significant.

Given the limited statistics, a further analysis beyond the line
detection is not warranted. However, given our stacking method, which
is biased for sources that are bright and/or with a long exposure, it
is at least worth examining whether Arp~220 and NGC~3690{\ts}E, the
two known Fe{\ts}K emitters from their XMM-Newton spectra, dominate the line
detection. These two objects and the rest of the HXQ sample were
integrated separately, and their spectra are presented in Fig.~2b and
2c. Both spectra show a $\simeq 2\sigma$ excess in the Fe{\ts}{\sc
  xxv} band with comparable line fluxes. This means that, besides
Arp~220 and NGC~3690{\ts}E, there is a significant contribution by
other galaxies to the detected line flux. The higher continuum level
in the spectrum of the sample without Arp~220 and NGC~3690{\ts}E
(Fig.~2c) suggests that some sources without strong Fe{\ts}K emission
are present, and that they may be dominated by HMXBs as found in
nearby star-forming galaxies.

Fitting a power-law to the continuum gives a photon index of $\Gamma =
1.1\pm 0.5$, indicating a rather hard conitnuum above 4{\ts}keV. The
equivalent width of the excess in the Fe{\ts}K band with respect to
the power law continuum is $EW = 0.9\pm 0.3$ keV. The 90\% upper limit
of the EW for a 6.4{\ts}keV line is 120{\ts}eV.

Fitting a thermal spectrum (MEKAL) gives a temperature of $kT =
8\pm2${\ts}keV and an Fe metallicity consistent with the solar
value. Considering that the integrated spectrum is diluted by some
sources that contribute only to the continuum emission (see above),
the Fe metallicity of the line emitting sources is likely to be
super-solar.

Regardless of the choice of spectral model, the strong high-ionization
Fe{\ts}K line is not compatible with the X-ray spectrum of known
HMXBs. Fe{\ts}K emission is often observed from HMXBs in our Galaxy,
but the major component is the cold line at 6.4{\ts}keV and its EW is
on average $\sim$200-300{\ts}eV (e.g. White et al. 1983). The much
stronger, high-ionization line in the integrated spectrum of the HXQ
sample suggests that X-ray binaries are not the primary source of the
4-8{\ts}keV emission.

\section{Discussion}

The general quietness of hard X-ray emission in the HXQ galaxies
(Fig. 1) is likely related to the detection of the high-ionization
Fe{\ts}K feature, which rules out HMXBs as the primary origin of the
faint hard X-ray emission.

The powerful FIR emission in the HXQ (U)LIRGs predicts an abundance of
HMXBs if star formation is the dominat power source. The Chandra
spectra, however, suggests these sources contribute little to the hard
X-ray emission. Therefore, either the bulk of the HMXBs are missing or
they are heavily obscured from view. Suppose heavy obscuration is the
cause. Since the Fe{\ts}K band is still transparent to \nH $\sim
10^{24}$ \psqcm, the obscuring column has to be larger, perhaps the
order of $10^{25}$ \psqcm. A dense, nuclear molecular disk (with a
dynamical mass of the order of $10^9 M_{\odot}$) is often found in the
central part of ULIRGs (e.g., Bryant \& Scoville 1999) and thus, such
a large column may not be unrealistic. The size of the nuclear
molecular disk is generally found to be a few 100 pc but the densest
part can be as small as 30 pc, as found by high-resolution studies of
Arp 220 (Downes \& Eckart 2007, Sakamoto et al.  2008, Aalto et
al. 2009), where a rough estimate of the density is $10^5$-$10^6$
cm$^{-3}$. According to Grimm et al. (2003), the number of HMXBs with
$L_{\rm X} > 2\times 10^{38}$ \ergps, which would dominate the
integrated luminosity of the entire HMXB population, is $\sim 300
(SFR/100 M_{\odot}$ yr$^{-1}$). For $SFR=200 M_{\odot}$ yr$^{-1}$, as
estimated for Arp 220, the number is $\sim 600$, and most of them need
to be confined within 30{\ts}pc in order for their radiation to be
suppressed. The required stellar density is unusually high and could
be a problem in this particular case. The other galaxies may have a
larger starburst region and could avoid this problem provided the
required mass of the obscuring gas does not exceed the dynamical mass.

Once the HMXBs are obscured from view, the Fe{\ts}{\sc xxv} spectrum
can be explained by high-temperature ($T\sim 10^8${\ts}K), thermal gas
produced by a starburst. Possible sources are (i) an internally
shocked hot bubble\footnote{This should be distinguished from a much more
  extended, soft X-ray nebula, which most likely originate from the
  swept-up interstellar medium (e.g., Tomisaka \& Ikeuchi 1988).}
produced by thermalizing the energy of supernovae (SNe) and stellar
winds, as predicted by, e.g., Chevalier \& Clegg (1985); and (ii)
collective luminous SNe. The latter is ruled out if HMXBs are all
embedded in heavy obscuration because the SNe should also be embedded
in the same obscuring material. The former would excavate the
obscuration and become visible. With the assumed high SFR, the
luminosity and the spectrum with strong Fe{\ts}{\sc xxv} can be
reproduced (e.g., Iwasawa et al. 2005 for Arp 220). The diffuse hard
X-ray emission seen in M~82 (Griffiths et al. 2000, Strickland \&
Heckman 2007, Ranalli et al. 2008), after removing resolved X-ray
binaries, has, in fact, a comparable X-ray to FIR ratio with that of
our GOALS HXQ sample. The strong Fe line is also well matched with the
prediction for metal enriched gas of this origin.

With the presence of dense nuclear gas, a heavily obscured AGN is also
a possible explanation for the X-ray spectra of our HXQ sample.
Massive black holes in the process of rapid growth are naturally expected 
to be present in (U)LIRGs. When a central AGN is deeply buried with a
covering factor nearly unity, reprocessed light from the obscuring
matter would have difficulty escaping. The high-ionization Fe{\ts}K line
could then originate from extended, low density gas which is photoionized
by the AGN. In a number of Compton thick Seyfert 2 galaxies, e.g.,
NGC~1068, Fe{\ts}{\sc xxv} has been seen (but normally weaker than the cold
line at 6.4{\ts}keV). This is in fact the feature originally predicted by
Krolik \& Kallman (1987) when assuming a likely scattering medium to
produce the polarized broad-line region in NGC~1068 (Antonucci \&
Miller 1985). Where the cold X-ray reflection is suppressed, the
weak, highly photoionised gas could be the only observable feature. We
note that Fe{\ts}{\sc xxv} has been observed as the primary Fe{\ts}K feature
in some Compton-thick AGN residing in the ULIRGs IRAS F00183--7113
(Nandra \& Iwasawa 2007, Ruiz et al. 2007), The Superantennae (Braito et al. 2009), and possibly UGC~5101.

In Arp 220, 
recent measurements of a compact dust emission source
with a steep temperature gradient in the western nucleus provides an
argument for an AGN (Downes \& Eckart 2007, Aalto et al. 2009; but see
Sakamoto et al. 2008). Millimeter wavelength molecular line observations 
also suggest the presence of X-ray dominated chemistry, which favors 
heating by AGN (Aalto et al. 2007, Imanishi et al. 2007; 
e.g., Meijerink \& Spaans 2005 for theory). The similarity
of the X-ray properties suggests that the same explanation might apply
for at least some of the GOALS HXQ galaxies -- i.e., despite the lack of
outward evidence of an AGN, a significant fraction of the infrared output
could be powered by a heavily obscured AGN.

\section*{Acknowledgements}

This research was supported in part by NASA through {\it Chandra}
award Number GO7-8108A, issued by the {\it Chandra X-Ray Observatory},
which is operated by the Smithsonian Astrophysical Observatory for and
on behalf of NASA under contract NAS8-39073. 
We acknowledge use of the NASA/IPAC
Extragalactic Database (NED), and the software packages CIAO and HEASoft.

\end{document}